\documentclass[runningheads]{llncs}
\usepackage[T1]{fontenc}
\usepackage{graphicx}
\usepackage{amssymb}
\usepackage{booktabs}
\usepackage{listings}
\usepackage{multirow}
\usepackage{arydshln}
\usepackage{hyperref}
\usepackage{bbding}

\begin{document}

%
\title{Statically Typed Purely Functional Definitional Interpreters for {\it Essentials of Programming Languages}}
\titlerunning{Statically Typed Purely Functional Definitional Interpreters for {\it EOPL}}
%
\author{Kwanghoon Choi\orcidID{0000-0003-3519-3650} \Envelope}
\authorrunning{Kwanghoon Choi}
%
\institute{Chonnam National University, Gwangju 61186, Republic of Korea
\email{kwanghoon.choi@jnu.ac.kr}}
\maketitle              
\begin{abstract}
Definitional interpreters provide an executable and pedagogically effective way to explain programming language semantics. However, when implemented in a dynamically typed and effectful metalanguage, semantic components such as stores, continuations, handlers, substitutions, and scheduling states may remain implicit in the host-language implementation, making it harder to see how language extensions affect interpreter structure. This paper reconstructs the family of interpreters from \textit{Essentials of Programming Languages} (EOPL) in Haskell, a statically typed and purely functional metalanguage, for pedagogical use. We preserve the object languages, examples, and tests of EOPL while making their semantic components explicit through algebraic data types, evaluator signatures, and explicit semantic domains. Across 15 interpreters and 6 type checkers, language extensions are exposed as systematic type-directed changes.

The reconstruction is currently used in undergraduate and graduate programming language courses, where students learn interpreter construction through incremental and test-driven development. We further illustrate its pedagogical use through three case studies that progressively extend students' experience: composing mutable state and exception handling, deriving an actor-based interpreter from the typed structure of the EOPL thread interpreter, and translating the CHECKED interpreter and type checker into Rocq to mechanize type safety. Together, the classroom experience and case studies suggest that the reconstruction provides a concrete bridge from learning programming language concepts through executable interpreters to language design and mechanized reasoning.
\end{abstract}
\section{Introduction}
\label{sec:intro}

Programming language textbooks written 
so far can be broadly categorized into two groups. The first group focuses on the taxonomy 
of programming languages, classifying programming paradigms such as functional languages 
(e.g., Scheme, ML, Haskell), object-oriented languages (e.g., C++, Java, Python), and logic languages 
(e.g., Prolog), and introducing the key features of representative programming languages 
in each paradigm.

The second group deals with programming language semantics, focusing on rigorously defining 
and understanding the behavior of programs. It aims to provide the 
basic mathematical techniques necessary for those who are beginning a study of the semantics 
and logics of programming languages \cite{Winskel1993}. 
Some textbooks in the group are shown in Table \ref{tbl:plbooks}.
The table highlights how education in the second group is shifting to emphasize
not only the mathematical techniques for defining semantics but also the approach 
of explaining programming languages through {\it interpreters}.

\begin{table}[t]
\centering
\caption{A Brief History of Semantics-based Programming Language Textbooks}
\label{tbl:plbooks}
{\small
\begin{tabular}{lc}
\toprule
Title of Programming Language Textbooks (Year) & Languages\\ 
\midrule
The Formal Semantics of Programming Languages (1993) & Math \\ 
Theories of Programming Languages (1998) & Math \\ 
Types and Programming Languages (2002) & Math \& OCaml \\ 
Semantics with Applications: An Appetizer (1992-2007) & Math \\ 
Structure and Interpretation of Computer Programs (1985-2022) & Scheme/JavaScript \\
Essentials of Programming Languages (1983-2008) & Scheme \\ 
Programming Languages: Appl. \& Interpret. (2000-2023) & Scheme \\
Software Foundations (2011--2026) & Rocq \\ Programming Language Foundations in Agda (2019--2026) & Agda \\
\bottomrule
\end{tabular}
}
\end{table}

A more recent direction in this second group is to use interactive theorem provers as textbook languages. Textbooks such as \textit{Software Foundations} \cite{Pierce:SF2}
and \textit{Programming Language Foundations in Agda} \cite{plfa22.08} present programming language semantics not only as mathematical definitions or executable interpreters, but also as mechanized formal developments. In this approach, syntax, operational semantics, type systems, and meta-theoretic properties such as progress and preservation are encoded directly in a proof assistant, and their proofs are checked mechanically. Thus, this line of textbooks can be seen as combining the mathematical semantics tradition with executable and machine-checkable artifacts.

\begin{figure}[t]
\centering
\includegraphics[width=\textwidth]{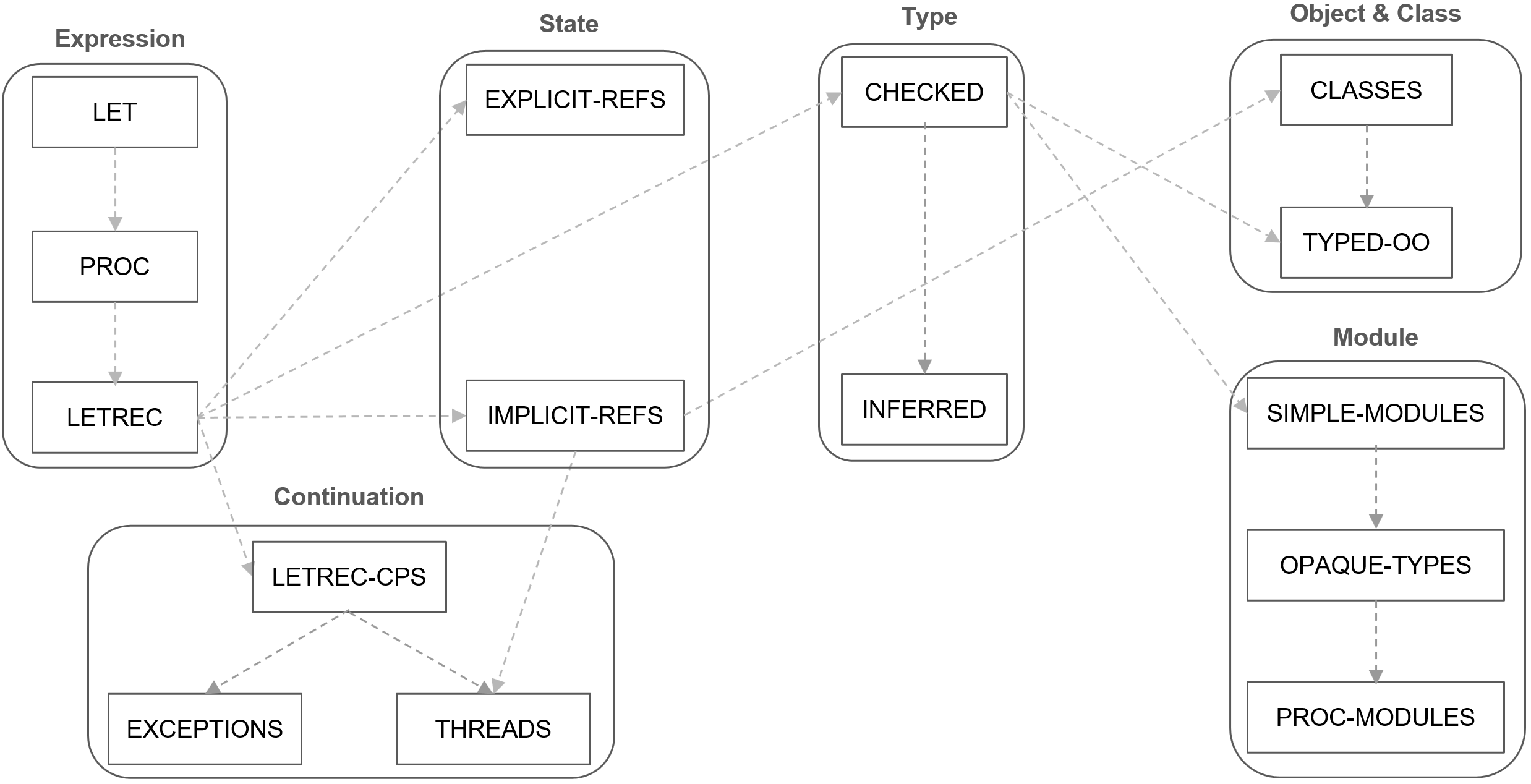}
\caption{Each box denotes a programming language in EOPL, and an arrow from language A to language B indicates that B is obtained by adding new features to A. Table \ref{tbl:pldescription} summarizes the main feature of each language.}
\label{fig:eoplinhaskell}    
\end{figure}

We favor this interpreter-based approach because 
providing a semantics model that can be directly executed through a working 
program makes it easier for students to engage with and learn the material. 
Furthermore, directly modifying this program can facilitate the design and exploration of new programming languages, 
making the process more intuitive and enjoyable through hands-on experimentation.
A textbook such as {\it Essentials of Programming Languages} ({\it EOPL}) \cite{eopl3rd} demonstrates this approach particularly well: a family of small languages is introduced incrementally, and each new language feature is explained by extending an interpreter for a previous language as shown in Figure \ref{fig:eoplinhaskell} 

Moreover, EOPL covers, within a single incremental interpreter-based framework, a particularly broad spectrum of programming language features, including the lambda calculus, recursion, explicit and implicit states, continuations, exceptions, concurrency, type checking, type inference, modules \& interfaces, opaque types, module procedures, and classes \& objects, as summarized in Table \ref{tbl:pldescription}.
To the best of our knowledge, we have not found another textbook or related work that covers this breadth of language features within a single incremental interpreter-based framework.

\begin{table}[ht]
    \centering
    \caption{The Status of the 15 Programming Languages in EOPL (IN: Interpreter, TC: Type Checker)}
    {\small
    \begin{tabular}{lccl}
        \toprule
        \textbf{Topics/PLs} & \textbf{\ \ IN \ \ } & \textbf{\ \ TC\ \ } & \textbf{Description} \\
        \midrule
        \multicolumn{4}{l}{\textbf{Expression}} \\
        \texttt{LET} & $\checkmark$ & n/a & Lexical scope \\
        \texttt{PROC} & $\checkmark$ & n/a & The $\lambda$-calculus \\
        \texttt{LETREC} & $\checkmark$ & n/a & Recursive Procedure \\
        \midrule
        \multicolumn{4}{l}{\textbf{State}} \\
        \texttt{EXPLICIT-REFS} & $\checkmark$ & n/a & C-style Pointer \\
        \texttt{IMPLICIT-REFS} & $\checkmark$ & n/a & Java-style Reference \\
        \midrule
        \multicolumn{4}{l}{\textbf{Continuation}} \\
        \texttt{LETREC-CPS} & $\checkmark$ & n/a & Continuation-Passing Style \\
        \texttt{EXCEPTIONS} & $\checkmark$ & n/a & Exception Handling \\
        \texttt{THREADS} & $\checkmark$ & n/a & Multi-threads and Mutexes \\
        \midrule
        \multicolumn{4}{l}{\textbf{Types}} \\
        \texttt{CHECKED} & $\checkmark$ & $\checkmark$ & Type Checking \\
        \texttt{INFERRED} & $\checkmark$ & $\checkmark$ & Type Inference \\
        \midrule
        \multicolumn{4}{l}{\textbf{Modules}} \\
        \texttt{SIMPLE-MODULES} & $\checkmark$ & $\checkmark$ & Simple Modules and Interfaces \\
        \texttt{OPAQUE-TYPES} & $\checkmark$ & $\checkmark$ & Opaque and Transparent types \\
        \texttt{PROC-MODULES} & $\checkmark$ & $\checkmark$ & Module Procedures \\
        \midrule
        \multicolumn{4}{l}{\textbf{Objects \& Classes}} \\
        \texttt{CLASSES} & $\checkmark$ & n/a & Class, Object, Interface, Inheritance \\
        \texttt{TYPED-OO} & $\checkmark$ & $\checkmark$ & Type Checking OO-PL \\
        \bottomrule
    \end{tabular}
    }
    \label{tbl:pldescription}
\end{table}


The interpreter-based approach requires selecting a meta-language for concrete implementation. 
EOPL asserts that any language supporting both first-class procedures and assignments
(ML, Common Lisp, Python, Ruby, etc.) is suitable for implementing interpreters.
Among them, it chooses Scheme—a dynamically typed functional language with support for assignments \cite{eopl3rd}.

However, the essence of EOPL could be effectively taught using types, making the explanation of interpreters more direct and intuitive. 
A well-known advantage of static types is that function type signatures effectively describe the function's behavior.
Static types for interpreter functions can also highlight similarities and differences among interpreters in a systematic way. 
In fact, the third edition of EOPL newly incorporates informal type declarations, called contracts, 
in nontrivial function definitions to clarify key implementation ideas, 
which aligns well with our motivation. 
However, these contracts do not fully capture all essential aspects of the implementations, leaving certain details implicit.
Then, a thought arose: why not use a statically typed functional language for EOPL instead,
making these aspects explicit?

Adopting a purely functional language enhanced our strategy of using static types in EOPL as
every side effect such as state and error handling in the interpreters must be explicitly reflected in their
function type signatures, which would otherwise remain hidden under implicit effects. 

Our goal is to develop statically typed and purely functional EOPL-style definitional interpreters. 
We do not propose new object languages, nor do we seek to replace the original textbook semantics. 
Instead, we intentionally preserve the object languages, examples, and tests of EOPL as a reference point, 
while reconstructing their implementation architecture in Haskell. This allows us to separate two concerns: 
the object-language behavior remains close to the original EOPL interpreters, whereas the metalanguage representation 
exposes the semantic components that were previously implicit.

The reconstruction is currently used in two programming language courses at the author's university, 
targeting different academic levels. The earlier part of EOPL has been taught to third-year undergraduate students, 
while the later part has been taught to first-year master’s students. The corresponding lecture materials and 
Haskell source code are publicly available\footnote{\url{https://github.com/kwanghoon/Lecture_EOPL}}. 


To further illustrate its pedagogical use, we present three case studies that expose students to progressively broader uses 
of typed definitional interpreters: combining mutable state and exception handling demonstrates how independently introduced 
language features can be composed; deriving an actor-based interpreter from the typed structure of \texttt{THREADS} demonstrates 
how an existing interpreter can guide the design of a new language; and translating the \texttt{CHECKED} interpreter and 
type checker into Rocq demonstrates how executable interpreters can serve as a starting point for mechanized reasoning. 
Together, the classroom use and these case studies show how the reconstruction can support a progression from understanding 
language features, through interpreter extension and language design, to formal verification.

In summary, this paper contributes a statically typed and purely functional reconstruction of the EOPL interpreter family, 
together with a typed design space that makes the semantic structure of language extensions explicit. The reconstruction is 
intended for pedagogical use and is supported by classroom experience, accompanying teaching materials, and case studies 
on feature composition, language extension, and mechanized reasoning.

Section \ref{sec:background} reviews EOPL. Section \ref{sec:design} then presents a typed design space and 
a type-directed reconstruction of EOPL in a pure metalanguage. Section \ref{sec:casestudies} presents three case studies 
that further illustrate the pedagogical use of our EOPL reconstruction. Section \ref{sec:discussion} discusses our teaching experience with the reconstruction and related work, and Section \ref{sec:conclusion} concludes.

\section{Background: EOPL-style interpreters and type checkers}
\label{sec:background}

The {\it Essentials of Programming Languages (EOPL)} is a well-known textbook that presents a family of small programming languages organized around six major topics: expressions, state, continuations, types, modules, and objects\&classes. As shown in Figure \ref{fig:eoplinhaskell}, the textbook introduces 15 languages incrementally: each box denotes a programming language, and an arrow from language A to language B indicates that B is obtained by adding new features to A.


Table \ref{tbl:pldescription} summarizes the main feature of each language.
%
The expression languages \texttt{LET}, \texttt{PROC}, and \texttt{LETREC} introduce the basic structure of definitional interpreters for lexical binding, procedures, and recursive procedures. The state languages explain two styles of mutable state: \texttt{EXPLICIT-REFS} models C-style references, where locations are explicitly manipulated by programs, whereas \texttt{IMPLICIT-REFS} models Java-style references, where locations are used implicitly. The continuation languages introduce continuations as explicit representations of control contexts and use them to explain continuation-passing style, exception handling, and threads. The typed languages present both type checking and type inference. The module languages introduce mechanisms for programming in the large, and the object\&class languages present object-oriented features as an alternative form of program organization.

In EOPL, all 15 languages are presented through executable interpreters, and the later languages in the topics of types, modules, and objects\&classes are also accompanied by type checkers. In our reconstruction, we implemented these interpreters and type checkers as working Haskell programs. The purpose of this reconstruction is not to introduce new object languages, but to expose the implementation architecture of the EOPL interpreter family in a statically typed and purely functional metalanguage. In the following sections, we use this artifact to analyze how language extensions correspond to systematic changes in evaluator signatures, semantic domains, and explicit runtime components.

\section{A Typed Design Space and Type-Directed Reconstruction of EOPL}
\label{sec:design}

\subsection{Running Example: LET}

Before discussing the full typed design space, we introduce the basic semantic domains used throughout this paper with a small example from the LET language in Figure \ref{fig:letastev}. Consider the following LET program:
\[
\texttt{let x = 3 in -(x,1)}
\]
This program is written in the concrete syntax of the object language. After parsing, it is no longer treated as a string, but as an abstract syntax tree. In our Haskell representation, it is represented by an expression of type \texttt{Exp}, for example:
\[
\begin{array}{l}
\texttt{Let\_Exp "x"}
\quad \texttt{(Const\_Exp 3)}
\quad \texttt{(Diff\_Exp (Var\_Exp "x") (Const\_Exp 1))}
\end{array}
\]
Thus, \texttt{Exp} represents the abstract syntax of object-language programs. It is not the result of executing a program; rather, it describes the syntactic structure of a program phrase before evaluation.

Evaluation takes place relative to an environment. In the example above, the body \texttt{-(x,1)} must be evaluated under the information that \texttt{x} is bound to the value of \texttt{3}. This information is represented by an environment of type \texttt{Env}, such as:
\[
\texttt{Extend\_env "x" (Num\_Val 3) Empty\_env}
\]
An environment is therefore a semantic context that associates variable names with their meanings. In the basic LET interpreter, this meaning is an expressed value. The evaluator consequently has the type:
\[
\texttt{value\_of :: Exp -> Env -> ExpVal}
\]
This type can be read as follows: given an expression and an environment, the interpreter produces an expressed value.

The result of evaluating the example program is the integer value \texttt{2}. However, in the interpreter, this result is not represented directly as a Haskell integer of type \texttt{Int}. Instead, it is represented as an object-language value:
\[
\texttt{Num\_Val 2}
\]
The type of this value is \texttt{ExpVal}. Thus, \texttt{ExpVal} represents the runtime values of the object language. In LET, expressed values include numbers and booleans. As the language is extended, the definition of \texttt{ExpVal} is also extended: procedure values are added in PROC and LETREC, reference values are added in EXPLICIT-REFS, mutexes and queues are added in THREADS, and object values are added in CLASSES.

For typed languages, one more distinction is necessary as follows. The expression \texttt{Const\_Exp 3} is an abstract syntax tree of type \texttt{Exp}; when evaluated, it produces the expressed value \texttt{Num\_Val 3} of type \texttt{ExpVal}; when statically analyzed, it is assigned the object-language type \texttt{TyInt} of type \texttt{Type}. These three levels should not be confused:
\[
\texttt{Const\_Exp 3 :: Exp}
\qquad
\texttt{Num\_Val 3 :: ExpVal}
\qquad
\texttt{TyInt :: Type}
\]
The first is a syntactic representation of a program phrase, the second is a runtime result, and the third is a static classification.

This simple example explains the core domains that appear repeatedly in the rest of the paper. The basic evaluator for LET has the form \texttt{Exp -> Env -> ExpVal}. Later language features modify this structure in systematic ways. Mutable state adds an explicit \texttt{Store}; control features add continuations of type \texttt{Cont}; concurrency adds scheduling state; and type checking and type inference introduce type environments, types, substitutions, and error effects. The main claim of this paper is that these changes become visible and systematic when EOPL-style interpreters are reconstructed in a statically typed and purely functional metalanguage.

\begin{figure}[ht]
\begin{center}
\begin{minipage}[t]{.48\linewidth}
{\small
\begin{lstlisting}[mathescape=true,basicstyle=\ttfamily]
data Exp =
    Const_Exp  Int
  | Diff_Exp   Exp Exp
  | IsZero_Exp Exp
  | If_Exp     Exp Exp Exp
  | Var_Exp    Ident
  | Let_Exp    Ident Exp Exp
\end{lstlisting}
}
\end{minipage}
\begin{minipage}[t]{.48\linewidth}
{\small 
\begin{lstlisting}[mathescape=true,basicstyle=\ttfamily]
data ExpVal =
    Num_Val  Int
  | Bool_Val Bool

data Env = Empty_env
  | Extend_env Ident ExpVal Env
\end{lstlisting}
}
\end{minipage}
\end{center}
\caption{Abstract Syntax Trees (\texttt{Exp}), Expressed Values (\texttt{ExpVal}), and Environments (\texttt{Env}) for \texttt{LET}, where \texttt{Ident} a type synonym for \texttt{String}}
\label{fig:letastev}
\end{figure}

\subsection{A Reconstruction of the EOPL interpreters and type checkers}
We now explain how static types help structure and clarify the elements of programming languages in our reconstruction of EOPL. Algebraic data types provide a natural way to represent programming language constructs by composing sum types, product types, and recursive types \cite{burstall1980}. This allows language elements to be expressed as mathematical domains ({\it inductive sets}), where each element---such as expressions (\texttt{Exp}), environments (\texttt{Env}), continuations (\texttt{Cont}), stores (\texttt{Store}), expressed values (\texttt{ExpVal}), and types (\texttt{Type})---is represented by a corresponding algebraic data type, as summarized in Table \ref{tbl:typed-design-space}.

\begin{table}[t]
\centering
\caption{Main axes of the typed design space}
\label{tbl:typed-design-space}
\begin{tabular}{l | l | l}
\hline
\textbf{Axis} & \textbf{Meaning} & \textbf{Example}\\
\hline
\texttt{Exp} 
  & Abstract syntax tree & \texttt{Const\_Exp 3}, \texttt{Var\_Exp "x"}, \texttt{Let\_Exp ...}\\
  
\texttt{Env} 
  & Variable/recursive bindings & x $\mapsto$ \texttt{Num\_Val 3}\\

\texttt{ExpVal} 
  & Semantic values & \texttt{Num\_Val 3}, \texttt{Bool\_Val True}, \texttt{Proc\_Val ...}\\

\texttt{Store} 
  & Mutable state & \texttt{location} $\mapsto$ \texttt{ExpVal} \\

\texttt{Cont} 
  & Control context & \texttt{Diff1\_Cont exp2 env cont}, \texttt{End\_Cont} \\

\texttt{Type}
  & Types, interfaces, classes & \texttt{TyInt}, \texttt{TyBool}, \texttt{TyFun ty1 ty2}, \texttt{TyClass Ident} \\

\hline
\end{tabular}
\end{table}

Static types are useful not only for representing these domains, but also for distinguishing elements that may look similar at the surface level. For example, a number in a program is initially represented as the string \texttt{"123"} of type \texttt{String}. After parsing, it is transformed into an abstract syntax tree node such as \texttt{Const\_Exp 123} of type \texttt{Exp}. An interpreter then evaluates this tree to produce the expressed value \texttt{Num\_Val 123} of type \texttt{ExpVal}. Finally, in a typed language, the static type of this expressed value is represented as \texttt{TyInt} of type \texttt{Type}. Thus, static types make explicit the different stages through which a program phrase passes: concrete syntax, abstract syntax, runtime value, and static type.

Beyond distinguishing individual language components, static types also reveal how the architecture of an interpreter changes as a language is extended. We therefore characterize the reconstruction of EOPL interpreters as a set of recurring type-directed extension patterns. Some language extensions preserve the overall evaluator type and only extend the syntactic and semantic domains, as in pure expression extensions that add new forms to \texttt{Exp} and new values to \texttt{ExpVal}. Other extensions require changes to specific semantic components: binding constructs change the structure of \texttt{Env}, mutable state turns a direct evaluator of type \texttt{Exp -> Env -> ExpVal} into a store-passing evaluator of type \texttt{Exp -> Env -> Store -> (ExpVal, Store)}, and control features introduce continuations by changing the evaluator into a continuation-passing form such as \texttt{Exp -> Env -> Cont -> Answer}. Concurrency further refines the answer type into a thread representation, where \texttt{Thread} explicitly carries both a store and a scheduling state. Similarly, type inference extends ordinary type checking by adding substitutions and their refinement to the type-checking process. These patterns show that language evolution in the EOPL family can be understood not only as code modification, but also as systematic changes in evaluator signatures and semantic domains.

\subsubsection{The Structure of Interpreters}

The type signatures of various interpreters can illustrate how different language characteristics influence their structure,
as shown in Table \ref{tbl:interp_types}.
For \texttt{LET}, \texttt{PROC}, and \texttt{LETREC}, as well as \texttt{SIMPLE-MODULES}, \texttt{OPAQUE-TYPES}, and \texttt{PROC-MODULES},
the interpreter follows the fundamental type signature \texttt{Exp -> Env -> ExpVal}, 
where expressions are evaluated with an environment to produce a semantic value. 
Although \texttt{Exp}, \texttt{Env}, and \texttt{ExpVal} are defined differently across these languages to reflect variations in syntax, scoping and binding, and runtime values, respectively, these differences are abstracted away at the level of the evaluator type, yielding the common type signature.

\begin{table}[ht]
\caption{Types of Interpreters}
\label{tbl:interp_types}
{\small
\begin{tabular}{l  l}
\toprule
Interpreter & Type Signature \\
\midrule
\texttt{LET, PROC, LETREC} & \texttt{Exp -> Env -> ExpVal} \\  
\texttt{SIMPLE-MODULES}, & \\
\texttt{OPAQUE-TYPES}, & \\ 
\texttt{PROC-MODULES} & \\
\texttt{(CHECKED, INFERRED)} & \\ \hdashline[0.5pt/2pt]
\texttt{EXPLICIT-REFS}, & \texttt{Exp -> Env -> Store -> (ExpVal, Store)} \\ 
\texttt{IMPLICIT-REFS} & \\ \hdashline[0.5pt/2pt]
\texttt{CLASSES, TYPED-OO} & \texttt{ClassEnv -> Exp -> Env -> Store -> (ExpVal, Store)} \\  \hdashline[0.5pt/2pt]
\texttt{LETREC-CPS$^{(1)}$}, & \texttt{Exp -> Env -> Cont -> ExpVal} \\  
\texttt{EXCEPTIONS$^{(1),(2)}$}  & (1) \texttt{apply\_cont :: Cont -> ExpVal -> ExpVal} \\  
 & (2) \texttt{apply\_handler :: ExpVal -> Cont -> ExpVal} \\ \hdashline[0.5pt/2pt]
  & \texttt{type Thread = Store  -> SchedState -> (ExpVal, Store)} \\
\texttt{THREADS$^{(3)}$} & \texttt{Exp -> Env -> Cont -> Thread} \\  
 & (3) \texttt{apply\_cont :: Cont -> ExpVal -> Thread} \\  
\bottomrule
\end{tabular}
}
\end{table}

In \texttt{EXPLICIT-REFS} and \texttt{IMPLICIT-REFS}, an additional \texttt{Store} parameter is introduced
whereas the original EOPL textbook retains the same fundamental type signature thereby making stores implicit. 

\texttt{CLASSES} and \texttt{TYPED-OO}, inheriting the store-passing style interpreter from \texttt{IMPLICIT-REFS}, 
incorporate \texttt{ClassEnv} to support class-based object-oriented programming.

\texttt{LETREC-CPS} and \texttt{EXCEPTIONS} adopt continuation-style interpreters, while \texttt{THREADS} combines this approach with the store-passing style, incorporating \texttt{Cont} to express control contexts. 
\texttt{EXCEPTIONS} introduces two control flows: a normal execution using \texttt{apply\_cont} and an exceptional flow using \texttt{apply\_handler}.
\texttt{THREADS} supports concurrent execution with an arbitrary number of control flows, executing them one by one in a round-robin schedule. 
In \texttt{THREADS}, while threads in the original EOPL book are represented with a contract that corresponds to the type \texttt{() -> ExpVal}, 
our approach instead introduces \texttt{Store -> SchedState -> (ExpVal, Store)}, explicitly managing both store and scheduling state. 

Further details can be found in an extended report \cite{khchoi2026techreport}.

\subsubsection{The Structure of Type Checkers}

\begin{table}[ht]
\caption{Types of Type Checkers}
\label{tbl:typechecker_types}
{\small
\begin{tabular}{ll}
\toprule
Type Checker & Type Signature \\
\midrule
\texttt{CHECKED}, &  \texttt{Exp -> TyEnv -> Either ErrMsg Type} \\  
\texttt{SIMPLE-MODULES} & \\ \hdashline[0.5pt/2pt]
\texttt{TYPED-OO} & \texttt{StaticClassEnv -> Exp -> TyEnv -> Either ErrMsg Type} \\  \hdashline[0.5pt/2pt]
\texttt{INFERRED} & \texttt{Exp -> TyEnv -> Subst -> Either\_ ErrMsg (Type, Subst)} \\   \hdashline[0.5pt/2pt]
\texttt{OPAQUE-TYPES}, & \texttt{Exp -> TyEnv -> Either\_ ErrMsg Type} \\  
\texttt{PROC-MODULES} & \\ 
\bottomrule
\end{tabular}
}
\end{table}

The type signatures of various type checkers also illustrate their fundamental structure and variations,
as shown in Table \ref{tbl:typechecker_types}.
The basic type-checking structure follows the signature \texttt{Exp -> TyEnv -> Either ErrMsg Type}, 
where an expression (\texttt{Exp}) and a type environment (\texttt{TyEnv}) are given as input,
returning either a successfully determined type (\texttt{Type}) or an error message (\texttt{ErrMsg}) if the expression is ill-typed.
This structure is adopted by the type checkers in \texttt{CHECKED} and \texttt{SIMPLE-MODULES}.
\texttt{TYPED-OO} introduces a static class environment (\texttt{StaticClassEnv}) to the basic signature,
ensuring that type checking considers class hierarchies.

\texttt{INFERRED} allows programmers to omit type declarations in their programs, with a type inference system automatically determining 
and filling in the missing types.  The system extends the type-checking process by generating fresh type variables for missing types
and solving type equations constrained by these variables. 
The type signature of the type inference in \texttt{INFERRED}
introduces a type substitution (\texttt{Subst}) parameter 
and replaces \texttt{Either} with \texttt{Either\_}, a variant that incorporates an internal counter state for fresh name generation. 

In \texttt{OPAQUE-TYPES} and \texttt{PROC-MODULES}, the type checkers handle opaque types and procedural modules using essentially the same type signature as in \texttt{CHECKED} and \texttt{SIMPLE-MODULES},
but using \texttt{Either\_} for matching bound module variables. 

\subsubsection{Validating the EOPL Reconstruction}

We implemented all interpreters and type checkers in Haskell, each comprising approximately 50--364 lines of code, while preserving the behavior of the original object languages presented in EOPL. To validate this preservation, we reused the complete set of original EOPL test cases for each language and confirmed that our Haskell implementations pass them. These tests cover the expected evaluation and type-checking behavior of the corresponding object languages, including representative programs for core language constructs and feature extensions such as procedures, references, exceptions, continuations, threads, modules, and type checking. Passing the original test suites gives evidence that our statically typed and purely functional reconstruction is faithful to the operational behavior of the EOPL implementations, while replacing the surrounding infrastructure—such as parsers, environments, stores, and auxiliary data structures—with typed Haskell modules.

Passing the original EOPL test suites does not guarantee that the two implementations are behaviorally identical. Stronger validation could be obtained by randomly generating additional object-language programs and comparing their results across the Scheme and Haskell implementations, or by more directly comparing the corresponding implementations and their semantic components.


\section{Case Studies}
\label{sec:casestudies}

We present three case studies that illustrate progressively broader pedagogical uses of the reconstruction. First, we combine mutable state and exceptions by composing their type-level effects: \texttt{Store} for state, and \texttt{Cont} with handler continuations for exceptions. Second, we derive an actor-language interpreter from the typed structure of \texttt{THREADS}, showing how an existing interpreter can guide the design of a new language. Third, we translate the \texttt{CHECKED} interpreter and type checker into Rocq and prove type safety, showing how executable interpreters can serve as a starting point for mechanized reasoning. Together, these studies illustrate a progression from understanding language-feature composition, through language design, to formal verification.

\subsection{Combining State and Exceptions by Composing Evaluator Types}
This case study focuses on constructing an interpreter for a language that combines mutable state with exception handling. In the EOPL language family, mutable state is introduced by the reference languages, while exception handling is introduced through a continuation-based interpreter. Each feature changes the evaluator architecture in a different way: state requires the evaluator to thread a store, whereas exceptions require the evaluator to make control contexts explicit.

This combination represents a more involved instance of a type-directed reconstruction pattern. A stateful interpreter changes a direct evaluator of type
\texttt{Exp -> Env -> ExpVal}
into a store-passing evaluator of type
\texttt{Exp -> Env -> Store -> (ExpVal, Store)}.
On the other hand, an exception-handling interpreter introduces continuations and handlers, leading to a continuation-passing evaluator such as
\texttt{Exp -> Env -> Cont -> Answer}.
To combine the two features, the answer type must also account for mutable state. Thus, the reconstructed evaluator has the form
\texttt{Exp -> Env -> Cont -> Store -> (ExpVal, Store)}
or, equivalently, by defining
\texttt{Answer = Store -> (ExpVal, Store)},
the evaluator can be written as
\texttt{Exp -> Env -> Cont -> Answer}.

The important point is that state and exceptions do not merely add new expression forms to the language. They introduce distinct semantic components that must be reflected in the interpreter architecture. The store represents the mutable state of the computation, while the continuation and exception handler represent the normal and exceptional control flows. In the combined interpreter, both normal evaluation and exception propagation must preserve the current store explicitly.

Implementing this interpreter therefore requires coordinated modifications across several semantic components: the abstract syntax is extended with reference operations and exception constructs, \texttt{ExpVal} is extended with reference values when locations are explicit, the evaluator is transformed into a store-passing and continuation-passing form, and the continuation structure is extended with frames for stateful operations and exception handlers. Regression tests from the state and exception languages can be reused to check that each feature still behaves as intended, while additional tests are needed to check their interaction. 
We therefore added three new tests: one checking that an assignment performed before an exception is raised is visible to the handler, one checking that the handler can update an implicitly allocated variable, and one checking that the raised value is consistent with the store observed by the handler. 
Both the original regression tests and these three new interaction tests pass. The following is one representative test case, whose expected result is \texttt{0}.

\begin{center}
\begin{minipage}[t]{.8\linewidth}
{\small
\begin{lstlisting}[mathescape=true,basicstyle=\ttfamily]
let x = 5 in 
 try begin  set x = -(x, 1);  raise x   end
 catch (e) -(e, x)
\end{lstlisting}
}
\end{minipage}
\end{center}

This example illustrates a central benefit of the typed reconstruction. The combination of language features can be understood as the composition of their type-level effects: mutable state contributes an explicit \texttt{Store}, while exceptions contribute explicit control contexts through \texttt{Cont} and handler continuations. The resulting evaluator type makes the interaction between state and exceptional control flow visible, rather than hiding it in the mutable or dynamically scoped mechanisms of the metalanguage.

\subsection{Deriving \texttt{ACTORS} from the Type Structure of \texttt{THREADS}}
\label{sec:actors}

\begin{figure}[h]
\centering
\includegraphics[width=\textwidth]{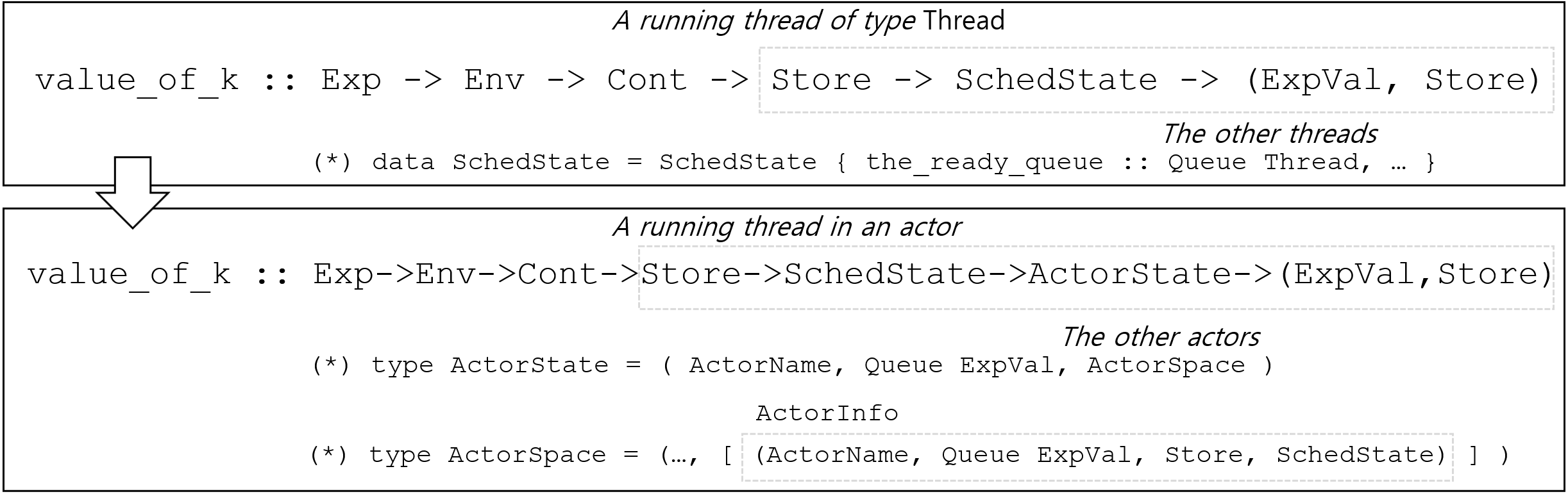}
\caption{Designing an Interpreter for \texttt{ACTORS} from the Interpreter for \texttt{THREADS}}
\label{fig:designactorsfromthreads}
\end{figure}

The second case study considers the design and implementation of an interpreter for a simple actor-based language, as described in Chapter 4 of {\it Programming Distributed Computing Systems: A Foundational Approach} \cite{varela2013}. This language extends the lambda calculus with actors. Each actor has a queue for receiving messages from other actors, a local store, and three actor operations: {\it send}, {\it ready}, and {\it new}. A Ping-Pong message exchange example written in \texttt{ACTORS} is shown in Figure \ref{fig:actorsexample}.

\begin{figure}[t]
\begin{center}
\begin{minipage}[t]{\linewidth}
{\small
\begin{lstlisting}[mathescape=true,basicstyle=\ttfamily]
let master = 
      proc(self)
	let pingServer = 
	      proc(self)
		letrec pingServer_self (msg) =
		   let PingMsg = 1 in
		   let PongMsg = 2 in
		   if zero? (-(msg, PingMsg))
		   then ready ( proc (masterSelf)
			    send (masterSelf, PongMsg, self))
		    else ready (pingServer_self)
		in ready(pingServer_self)
	in
        letrec master_self (msg) =
            let StartMsg = 0 in
            let PingMsg = 1 in
            let PongMsg = 2 in
            if zero? (-(msg,StartMsg))
            then
              let p = new(pingServer) in
                begin
                  send (p, PingMsg, self);
                  ready (master_self)
                end
            else if zero? (-(msg, PongMsg)) then print(400)
                 else ready (master_self)
        in ready(master_self)
in let StartMsg = 0 in
   let m = new (master) in 
       send (m, StartMsg)
\end{lstlisting}
}
\end{minipage}
\end{center}
\caption{A Ping-pong Example in \texttt{ACTORS}}
\label{fig:actorsexample}
\end{figure}

\texttt{ACTORS} is not part of the original EOPL language family. We use it to examine whether the typed design space extracted from EOPL can guide the construction of a new interpreter. The starting point is the \texttt{THREADS} interpreter, whose evaluator has the type
\texttt{Exp -> Env -> Cont -> Thread}
where
\texttt{Thread = Store -> SchedState -> (ExpVal, Store)}.
This type makes explicit two runtime components of a running thread: the mutable store and the scheduling state. In the original Scheme-based style, such components may be represented using global or mutable host-language state. In our reconstruction, they are explicitly represented in the type of \texttt{Thread}.

The actor-based language requires an additional level of runtime structure. A running actor must maintain its own actor name, message queue, local store, and scheduling state, while the actor system must also keep track of the other actors. Guided by the type structure of \texttt{THREADS}, we introduce \texttt{ActorState} to maintain the running actor's name (\texttt{ActorName}), its message queue (\texttt{Queue ExpVal}), and the surrounding actor space. The actor space consists of actor information records, where each \texttt{ActorInfo} contains an actor name, a message queue, a store, and a scheduling state.

Thus, the \texttt{ACTORS} interpreter is obtained by refining the runtime structure of \texttt{THREADS}. The scheduler state of \texttt{THREADS} accounts for the other threads in the ready queue; in \texttt{ACTORS}, this idea is lifted to the actor level, where each actor carries its own local queue, store, and scheduler state. As shown in Figure \ref{fig:designactorsfromthreads}, the type of the \texttt{THREADS} interpreter directly suggests the additional runtime components required for actors.

Like the \texttt{THREADS} interpreter in EOPL, the \texttt{ACTORS} interpreter simulates concurrency by interleaving the continuations of multiple running computations. Building on this implementation, we were able to modify the interpreter with the Cloud Haskell (\texttt{distributed-haskell}) library \cite{epstein2011towards}, allowing actor programs such as the Ping--Pong example to run in a distributed computing environment.

This case study shows that the typed reconstruction goes beyond reimplementing the original EOPL interpreters in Haskell. By making stores and scheduling states explicit, it provides a reusable design pattern for an actor language outside the textbook family. The Ping-Pong example then checks the interaction of message passing, actor creation, and scheduling.

\subsection{Mechanizing Type Safety for \texttt{CHECKED} in Rocq}
The third case study examines whether the typed and purely functional reconstruction can support formal verification. The \texttt{CHECKED} language in EOPL consists of a \texttt{LETREC}-style interpreter together with a type checker for the same language. Its expected meta-theoretic property is type safety: if a program is accepted by the type checker, then executing the program with the interpreter should not produce a value inconsistent with the assigned type.


Rather than encoding this soundness relationship using advanced features of Haskell's type system, we chose to mechanize the type safety argument in Rocq. The Haskell reconstruction was useful for this purpose because the interpreter and type checker were already written in a statically typed and purely functional style. As a result, most of the Haskell definitions could be translated into Rocq definitions directly.
%
The only substantial additional transformation concerns recursion. The \texttt{CHECKED} interpreter may define computations that do not terminate, whereas Rocq requires recursive functions to be structurally terminating. We therefore introduced a fuel parameter to the interpreter: evaluation proceeds while fuel remains, and a computation that would otherwise diverge terminates by exhausting the fuel. Apart from this standard fuel-based modeling of potentially nonterminating evaluation, the Rocq development follows the Haskell implementation closely.


The top-level theorem states that type checking is sound with respect to the fuel-based interpreter:

\begin{flushleft}
{\small
\texttt{Theorem type\_sound\_value\_or\_gas :
  forall (step : nat) (exp : Exp) (ty : Ty),
    type\_of\_program exp = inr ty -> 
    (exists v, value\_of\_program step exp = EvalValue v /\textbackslash{} value\_has\_type v ty)
    \textbackslash/ value\_of\_program step exp = EvalGasExhausted.}
}
\end{flushleft}

In words, if the \texttt{CHECKED} type checker assigns type \texttt{ty} to an expression \texttt{exp}, 
then the fuel-based interpreter either produces a value of type \texttt{ty} or exhausts the available fuel. Fuel exhaustion represents computations that do not terminate within the given bound, including potentially diverging computations.

This case study highlights another benefit of the reconstruction. If the interpreter had remained closer to the original Scheme style, with dynamically typed data representations and mutable or global host-language state, translating it into Rocq would likely have required additional semantic restructuring. In contrast, the Haskell version already makes the relevant semantic domains explicit as algebraic data types and expresses computation through pure functions. This made the Rocq mechanization close to a one-to-one translation of the Haskell artifact.
This experience is consistent with prior work such as seL4 \cite{sel4sosp2009} and hs-to-coq \cite{hs2coqcpp2018}, which use typed functional programs as a basis for mechanized reasoning.

Although this case study mechanizes type safety only for \texttt{CHECKED}, the same methodology suggests a path for future work. The EOPL family also contains languages with type inference, modules, and typed objects, namely \texttt{INFERRED}, the three module languages, and \texttt{TYPED-OO}. We would like to investigate whether their Haskell reconstructions can be translated into Rocq in a similarly direct manner, and to identify what additional transformations are required beyond the fuel parameter used for \texttt{CHECKED}.

\section{Discussion and Related Work}
\label{sec:discussion}

\subsection{Teaching Experience with the EOPL Reconstruction}

The reconstruction is currently used in two programming language courses at the author's university, 
targeting different academic levels. The earlier part of EOPL (8 interpreters and 1 type checker from \texttt{LET} through \texttt{CHECKED} except \texttt{THREADS}) has been taught to third-year undergraduates, while the later part (7 interpreters and 5 type checkers covering \texttt{THREADS} and the languages from \texttt{INFERRED} to \texttt{TYPED-OO}) has been taught to first-year  master’s students. The corresponding lecture materials and 
Haskell source code are publicly available. 


\subsubsection{From Math to Haskell} 


As semantics textbooks typically begin with the basic mathematical techniques needed to describe programming languages, our courses begin with a short introduction to the metalanguage used in the reconstruction. Since most students have little or no prior experience with Haskell, this preparation is essential. We use Lecture 9 of Part II of the Haskell MOOC\footnote{\url{https://haskell.mooc.fi/part2\#lecture-9-recap-of-part-1}} to introduce the basic notions needed for the course: types, functions, and programming with recursion. The material is particularly useful for self-study because it includes exercises together with tests that allow students to check their solutions.

To keep this prerequisite as small as possible, the reconstructed EOPL deliberately uses only a minimal typed functional vocabulary: algebraic data types, recursive functions, and primitive types such as integers, booleans, and strings. This choice not only makes the semantic components introduced by each language feature explicit, but also lowers the barrier to learning Haskell as the metalanguage. 

In particular, the notion of inductive sets introduced in Chapter 1 of EOPL as a mathematical technique for describing syntactic and semantic domains is important, and it is recast using Haskell algebraic data types, whose elements are constructed with data constructors and decomposed by pattern matching. These concepts provide students with the main Haskell tools needed to understand and extend the reconstructed interpreters.

\subsubsection{Test-driven and Incremental Development}


Inspired by agile development practices in software engineering, we structure exercises in our reconstructed EOPL around test-driven and incremental development, allowing students to build interpreters step by step while continuously validating their progress. Although passing the test suite does not guarantee interpreter correctness, it provides immediate feedback on each incremental implementation step, making it particularly useful in an interpreter-based learning approach.

To help students become familiar with both interpreter structure and basic Haskell programming, we provide partially completed projects in which the main interpreter or type-checker functions are left blank\footnote{\url{https://github.com/kwanghoon/Lecture_EOPL_Exercise}}. For example, the \texttt{value\_of} function of the \texttt{LET} interpreter is already divided into cases for the six expression forms, but the body of each case is removed for students to complete. This exercise encourages students to reconstruct the interpreter themselves. They can then use the same test cases employed for validating our reconstruction to check their implementations.

We additionally encourage an incremental, test-driven construction process. Students first run the test suite with an incomplete interpreter and observe the failures. They then implement only the functionality required by the simplest test cases and gradually increase the number of passing tests. For example, a test containing only a constant expression requires implementing only the \texttt{Const\_Exp} case for the \texttt{LET} interpreter. A subsequent subtraction test exposes the missing \texttt{Diff\_Exp} case, which students then implement and verify before proceeding to the next construct. Repeating this process until all tests pass helps students become accustomed to the structure of the \texttt{LET} interpreter and provides a disciplined way to manage the increasing complexity of later interpreters.


\subsubsection{Student Feedback from an Undergraduate Course}

We report feedback from 58 students in the undergraduate programming languages course offered in Spring 2026. Responses were collected on a five-point Likert scale, where 1 indicated “strongly disagree” and 5 indicated “strongly agree.” Students expressed strong support for the incremental organization of the course---starting from \texttt{LET} and gradually adding language features through a series of interpreters---with an average rating of 4.3. They generally favored learning semantics through executable interpreters rather than through mathematical definitions alone, although some students preferred mathematical descriptions to be used as complementary material. The test-driven and incremental development approach also received an average rating of 4.3. When asked whether the type-directed organization of the interpreters helped them understand their structure, students gave an average rating of 4.2, indicating a positive assessment of the use of types in organizing interpreter designs. In particular, they responded positively to representing syntactic and semantic domains uniformly with algebraic data types.

The main challenge was learning Haskell as the metalanguage. Except for four students, all were encountering Haskell for the first time, and even those four had limited prior experience. Students reported that becoming comfortable with Haskell typically required approximately one to one and a half months, often until shortly before the midterm examination. 
Feedback on Haskell was therefore mixed. Students identified the initial learning curve as a disadvantage, but also recognized benefits such as concise syntax, pattern matching, algebraic data types, and the way types guide interpreter construction. 
We believe that our strategy of deliberately restricting the reconstruction to a minimal set of Haskell features may have helped reduce the effort required to learn the metalanguage.

Among the eight object languages covered in the undergraduate course, students reported the greatest difficulty with recursion in \texttt{LETREC}, followed by continuations in \texttt{LETREC-CPS} and explicit references in \texttt{EXPLICIT-REFS}. 

Overall, a recurring positive response was that the course provided students, who had previously mainly used programming languages, with a new perspective on how programming languages themselves can be designed and implemented.

\subsubsection{From Classroom Use to Language Design}


One graduate student who used the artifact in both courses won third place in the graduate category of the ACM Student Research Competition at SPLASH 2025 for designing and implementing a multi-tier actor programming language \cite{hyeonjin2025}. The project built on the ACTORS case study in Section \ref{sec:actors} and our prior work on a polymorphic RPC calculus \cite{choi2020scp}. This provides a concrete example of how the artifact can support students in extending the reconstructed interpreters toward the design and implementation of new programming languages.


\subsection{From Explicit Interpreter Structure to Structured Semantics}

Directly exposing semantic components in algebraic data types and evaluator signatures, as in our reconstruction, is one straightforward way to structure semantics, but several more systematic approaches have also been developed.
Action semantics \cite{mosses1996theory} structures language definitions by decomposing behavior into reusable semantic actions. Monadic semantics \cite{sheng1995,wadler1992} represents computational effects such as state, exceptions, input/output, and nondeterminism using monads, allowing interpreters to be organized by their effects rather than by explicit parameter passing. Algebraic effects and handlers \cite{gordon2009,matija2015} further separate effect operations from their interpretations, providing a more modular way to describe and combine language features. 
Garby et al. \cite{garbyhaskell2024} similarly represent effects explicitly in a purely functional setting and use these representations to derive semantics-preserving compilers by equational reasoning. Although our goal differs, both approaches exploit explicit semantic structure to support systematic reasoning.
Interaction trees \cite{xiapopl2020itree} use a coinductive variant of free monads to represent recursive and effectful computations as executable semantic objects that can subsequently be interpreted by handlers. 



Compared with these approaches, our reconstruction exposes semantic components more directly and with less abstraction. In this sense, our interpreters can be viewed as presenting the underlying semantic components in an “unfolded” form, before they are packaged by abstractions such as monads. We view this simplicity as pedagogically useful: students can directly observe how each language extension changes the evaluator type and introduces corresponding semantic components, such as stores, continuations, and scheduling states. After encountering these components explicitly, more structured techniques can be introduced as abstractions over them. In this sense, the reconstruction provides a concrete bridge from pedagogical definitional interpreters to more systematic approaches to structuring executable and formal semantics.

\section{Conclusion and Future Work}
\label{sec:conclusion}

This paper reconstructed the family of EOPL-style definitional interpreters in Haskell for pedagogical use, preserving the original object languages, examples, and tests while making semantic components explicit through algebraic data types, evaluator signatures, and purely functional state passing. Across 15 interpreters and 6 type checkers, the reconstruction shows students how language extensions correspond to systematic changes in semantic domains and evaluator types. Its use in undergraduate and graduate programming language courses and the three case studies illustrate a progression from understanding language features, through extending and designing interpreters, to formal reasoning about programming languages.

Future work includes further evaluating the pedagogical effectiveness of the reconstruction through more systematic classroom studies, extending the approach to additional language features beyond EOPL, and scaling the Rocq mechanization to the remaining typed languages. 

\begin{credits}
\subsubsection{\ackname}This work was supported by the IITP grant (IITP-2023-RS-2023-00256629), the NRF of Korea grant funded by the Korea government (RS-2025-24523420), and the KISA Information Security College Support Project.
\subsubsection{\discintname}
The author has no competing interests to declare.
\end{credits}

\bibliographystyle{splncs04}


\appendix

\section{A Parsing System and Abstract Syntax Trees (\texttt{Exp})}
\label{sec:parsingsystem}


EOPL provides the SLLGEN tool \cite{eopl3rd}, which is based on LL parsing and written directly in Scheme, 
while we used the YAPB tool \cite{jintack2020}, which is based on LR parsing and implemented in Haskell. 
A key advantage of using a parsing library that allows both the parser and the interpreter to be written 
in the same implementation language is that it eliminates the need for a separate language (e.g., YACC \cite{dragonbook}, Happy \cite{happy}) 
specifically for parser construction.

\begin{figure}[p]
\centering
        {\small
        \begin{lstlisting}[mathescape=true,basicstyle=\ttfamily]
lexerSpec :: LexerSpec Token IO ()
lexerSpec = LexerSpec
  {
    lexerSpecList = 
      [ ("[ \t\n]" , \_ -> return Nothing),
        ("[0-9]+"  , \_ -> return (Just INTEGER_NUMBER)),
        ("\\-"     , \_ -> return (Just SUB)),
        ("\\("     , \_ -> return (Just OPEN_PAREN)),
        ("\\)"     , \_ -> return (Just CLOSE_PAREN)),
        ("\\,"     , \_ -> return (Just COMMA)),
        ("zero\\?" , \_ -> return (Just ISZERO)),
        ("if"      , \_ -> return (Just IF)),
        ("then"    , \_ -> return (Just THEN)),
        ("else"    , \_ -> return (Just ELSE)),
        ("let"     , \_ -> return (Just LET)),
        ("in"      , \_ -> return (Just IN)),
        ("\\="     , \_ -> return (Just EQ)),
        ("[a-zA-Z][a-zA-Z0-9]*", \_ -> return (Just IDENTIFIER))
      ]
  }
        \end{lstlisting}
        }

\centering
        {\small
        \begin{lstlisting}[mathescape=false,basicstyle=\ttfamily]
parserSpec :: ParserSpec Token Exp IO ()
parserSpec = ParserSpec
  {
    parserSpecList =
    [
      rule "Expression' -> Expression" 
        (\rhs -> return $ get rhs 1), 
      rule "Expression -> integer_number" 
        (\rhs -> return $ Const_Exp (read (getText rhs 1) :: Int)),
      rule "Expression -> - integer_number" 
        (\rhs -> return $ Const_Exp (-(read (getText rhs 2) :: Int))),
      rule "Expression -> - ( Expression , Expression )" 
        (\rhs -> return $ Diff_Exp (get rhs 3) (get rhs 5)),
      rule "Expression -> zero? ( Expression )" 
        (\rhs -> return $ IsZero_Exp (get rhs 3)),
      rule "Expression -> if Expression then Expression else Expression" 
        (\rhs -> return $ If_Exp (get rhs 2) (get rhs 4) (get rhs 6)),
      rule "Expression -> identifier" 
        (\rhs -> return $ Var_Exp (getText rhs 1)),
      rule "Expression -> let identifier = Expression in Expression" 
        (\rhs -> return $ Let_Exp (getText rhs 2) (get rhs 4) (get rhs 6))
    ]
  }
        \end{lstlisting}
        }
    \caption{Lexer and Parser Specifications for \texttt{LET} (An Excerpt)}
    \label{fig:lexer_parser}
\end{figure}

The lexer and parser for the \texttt{LET} language in Figure \ref{fig:lexer_parser} 
define its syntactic structure and tokenization process. 
The lexer (\texttt{lexerSpec}) recognizes keywords (\texttt{if}, \texttt{then}, \texttt{else}, \texttt{let}, \texttt{in}), 
symbols (\texttt{-}, \texttt{(}, \texttt{)}, \texttt{,}, \texttt{=}), numbers, and identifiers, converting them into corresponding tokens. 
The parser (\texttt{parserSpec}) then defines the grammar rules for expressions, supporting constants, negation, arithmetic operations
(\texttt{Diff\_Exp}), conditional expressions (\texttt{If\_Exp}), variable references (\texttt{Var\_Exp}), and let bindings (\texttt{Let\_Exp}). 
By combining these components, the system transforms raw text input into an abstract syntax tree (AST) that can be interpreted or further processed.





\section {Type Definitions in the Interpreters}

\subsection{Environment Types}

Table \ref{tbl:env_definitions} summarizes the environment (\texttt{Env}) definitions used in various interpreters,
highlighting how they evolve to accommodate different language features.
In \texttt{LET} and \texttt{PROC}, the environment consists of an empty environment (\texttt{Empty\_env}) and an extension mapping an identifier 
to an expression value (\texttt{Extend\_env}). 
\texttt{LETREC} extends this structure by adding recursive bindings with \texttt{Extend\_env\_rec}. By modifying \texttt{apply\_env}, to create a procedure value (also called a closure) at the time of a call rather than at the time of declaration, 
this environment structure enables a purely functional implementation of recursive procedures. 
All subsequent languages after \texttt{LETREC} in Figure \ref{fig:eoplinhaskell} adopt this structure
as is or with modification to support mutually recursive definitions. 
\texttt{EXPLICIT-REFS} and \texttt{IMPLICIT-REFS} introduce references, storing either expressed values, allowing programmers to handle locations
like C pointers, or locations, interpreting variables as references like in Java.
The module languages, 
\texttt{SIMPLE-MODULES}, \texttt{OPAQUE-TYPES}, and \texttt{PROC-MODULES} further extend the environment 
with \texttt{Extend\_env\_with\_module}, enabling module-based scoping. 
Finally, \texttt{CLASSES} and \texttt{TYPED-OO} incorporate class-based features with \texttt{Extend\_env\_with\_self\_and\_super}, 
to support the \texttt{self} and \texttt{super} keywords, enabling class hierarchies and method resolution.

Thus these variations reflect how different language features influence the design of an interpreter’s environment.

\begin{table}[ht]
\centering
\caption{Environment Type Definitions in Different Interpreters}
\label{tbl:env_definitions}
{\small
\begin{tabular}{l l}
\toprule
Interpreter & Environment Definition \\
\midrule
\texttt{LET}, \texttt{PROC} & 
\begin{lstlisting}[mathescape=true,basicstyle=\ttfamily]
data Env = Empty_env
  | Extend_env Ident ExpVal Env
\end{lstlisting} \\ 
\midrule
\begin{minipage}{.2\linewidth}
\texttt{LETREC}, \\
\texttt{LETREC-CPS}, \\
\texttt{EXCEPTIONS} \\
\texttt{(CHECKED,}\\
\texttt{INFERRED)}
\end{minipage}
& 
\begin{lstlisting}[mathescape=true,basicstyle=\ttfamily]
data Env = Empty_env
  | Extend_env Ident ExpVal Env
  | Extend_env_rec Ident Ident Exp Env
\end{lstlisting} \\ 
\midrule
\begin{minipage}[t]{.2\linewidth}
\texttt{EXPLICIT-REFS}
\end{minipage}
& 
\begin{lstlisting}[mathescape=true,basicstyle=\ttfamily]
data Env = Empty_env
  | Extend_env Ident ExpVal Env
  | Extend_env_rec [(Ident, Ident, Exp)] Env
\end{lstlisting} \\ 
\midrule
\begin{minipage}[t]{.2\linewidth}
\texttt{IMPLICIT-REFS}, \texttt{THREADS}
\end{minipage}
& 
\begin{lstlisting}[mathescape=true,basicstyle=\ttfamily]
data Env = Empty_env
  | Extend_env Ident Location Env
  | Extend_env_rec [(Ident, Ident, Exp)] Env
\end{lstlisting} \\ 
\midrule
\begin{minipage}[t]{.2\linewidth}
\texttt{SIMPLE-MODULES}, \\
\texttt{OPAQUE-TYPES}, \\
\texttt{PROC-MODULES}
\end{minipage}
& 
\begin{lstlisting}[mathescape=true,basicstyle=\ttfamily]
data Env = Empty_env
  | Extend_env Ident ExpVal Env
  | Extend_env_rec Ident Ident Exp Env
  | Extend_env_with_module Ident TypedModule Env
\end{lstlisting}
\\  
\midrule
\begin{minipage}{.2\linewidth}
        \texttt{CLASSES}, \\ \texttt{TYPED-OO} 
        \end{minipage} & 
\begin{lstlisting}[mathescape=true,basicstyle=\ttfamily]
data Env = Empty_env
  | Extend_env [Ident] [Location] Env
  | Extend_env_rec [(Ident,[Ident],Exp)] Env
  | Extend_env_with_self_and_super Object Ident Env
\end{lstlisting} \\ 
\bottomrule
\end{tabular}
}
\end{table}


Note that \texttt{apply\_procedure} is a runtime library that retrieves the saved environment from
a procedure value (closure), extends it with a binding between the  procedure's parameter variable
and the provided argument, and evaluates the body within the extended environment using \texttt{value\_of}. 

\begin{center}
{\small
\begin{lstlisting}[mathescape=true,basicstyle=\ttfamily]
apply_procedure :: Proc -> ExpVal -> ExpVal
apply_procedure (Procedure var body env) arg = 
  value_of body (extend_env var arg env)
\end{lstlisting}
}
\end{center}

\subsection{Expressed Value Type Definitions}
\label{sec:expvaldefs}

Table \ref{tab:expval} presents different variations of the \texttt{ExpVal} (Expressed Values) definition used in various interpreters. 
The \texttt{LET} interpreter supports only numerical and boolean values, represented by \texttt{Num\_Val} and \texttt{Bool\_Val}. 
The \texttt{PROC}, \texttt{LETREC}, \texttt{IMPLICIT-REFS}, \texttt{LETREC-CPS}, \texttt{CHECKED}, \texttt{INFERRED}, \texttt{SIMPLE-MODULES}, 
\texttt{OPAQUE-TYPES}, and \texttt{PROC-MODULES} interpreters extend this by adding \texttt{Proc\_Val}, which represents procedures. 
The \texttt{EXPLICIT-REFS} interpreter further introduces \texttt{Ref\_Val}, which represents reference values to memory locations. 
The \texttt{EXCEPTIONS} interpreter adds support for lists through \texttt{List\_Val}, enabling structured data manipulation. 
The \texttt{THREADS} interpreter builds upon this by incorporating concurrency-related values, 
including \texttt{Mutex\_Val} for mutual exclusion mechanisms and \texttt{Queue\_Val} for thread queues. 
Finally, the \texttt{CLASSES} and \texttt{TYPED-OO} interpreters introduce object-oriented features with \texttt{Object\_Val} to represent objects and \texttt{Uninitialized\_Val} for uninitialized fields, supporting class-based programming paradigms. 
This progression of \texttt{ExpVal} definitions reflects how different interpreters extend the expressiveness 
of values to accommodate various computational paradigms.

Note that the inclusion of \texttt{Queue\_Val} as one of the variants of expressed values is somewhat inconsistent, 
as the \texttt{THREADS} interpreter never produces queue values from evaluating expressions in the language. 
However, this inclusion stems from the strategy used to implement mutex queues. A mutex consists of a boolean value indicating 
whether the critical region it protects is currently in use and a queue for threads waiting when the critical region is occupied. 
While an expression may evaluate to a mutex, the queue itself is not directly manipulated by programs. 
However, mutex queues are stored in the same store that holds expressed values, and the functions 
\texttt{newref :: Store -> ExpVal -> (Location,Store)}, 
\texttt{setref :: Store -> Location -> ExpVal -> Store}, and
\texttt{deref :: Store -> Location -> ExpVal} 
is used uniformly to create, store, and retrieve both expressed values and mutex queues in the store. 
As a result, \texttt{Queue\_Val} was included 
in \texttt{ExpVal} to maintain consistency in storage and retrieval operations.

Also note that a special expressed value \texttt{Uninitialized\_Val} should have been introduced to indicate uninitialized state
in \texttt{CLASSES} and \texttt{TYPED-OO}.

\begin{table}[ht]
    \centering
    {\small
    \begin{tabular}{l l}
        \toprule
        Interpreter & Expressed Value Definition \\
        \midrule
        \texttt{LET} &
        \begin{lstlisting}[mathescape=true,basicstyle=\ttfamily]
data ExpVal =
    Num_Val {expval_num :: Int}
  | Bool_Val {expval_bool :: Bool}
        \end{lstlisting} \\
        \midrule
        \begin{minipage}{.3\linewidth}
        \texttt{PROC}, \texttt{LETREC}, \texttt{IMPLICIT-REFS}, \\ \texttt{LETREC-CPS}, 
        \texttt{CHECKED}, \texttt{INFERRED}, \\ \texttt{SIMPLE-MODULES}, 
        \texttt{OPAQUE-TYPES}, \\ \texttt{PROC-MODULES}
        \end{minipage} &
        \begin{lstlisting}[mathescape=true,basicstyle=\ttfamily]
data ExpVal =
    Num_Val {expval_num :: Int}
  | Bool_Val {expval_bool :: Bool}
  | Proc_Val {expval_proc :: Proc}
        \end{lstlisting} \\
        \midrule
        \texttt{EXPLICIT-REFS} &
        \begin{lstlisting}[mathescape=true,basicstyle=\ttfamily]
data ExpVal =
    Num_Val {expval_num :: Int}
  | Bool_Val {expval_bool :: Bool}
  | Proc_Val {expval_proc :: Proc}
  | Ref_Val {expval_loc :: Location}
        \end{lstlisting} \\
        \midrule
        \texttt{EXCEPTIONS} &
        \begin{lstlisting}[mathescape=true,basicstyle=\ttfamily]
data ExpVal =
    Num_Val {expval_num :: Int}
  | Bool_Val {expval_bool :: Bool}
  | Proc_Val {expval_proc :: Proc}
  | List_Val {expval_list :: [ExpVal]}
        \end{lstlisting} \\
        \midrule
        \texttt{THREADS} &
        \begin{lstlisting}[mathescape=true,basicstyle=\ttfamily]
data ExpVal =
    Num_Val   {expval_num  :: Int}
  | Bool_Val  {expval_bool :: Bool}
  | Proc_Val  {expval_proc :: Proc}
  | List_Val  {expval_list :: [ExpVal]}
  | Mutex_Val {expval_mutex :: Mutex } 
  | Queue_Val {expval_queue :: Queue Thread}
        \end{lstlisting} \\
        \midrule
        \begin{minipage}{.3\linewidth}
        \texttt{CLASSES}, \\ \texttt{TYPED-OO} 
        \end{minipage}
        &
        \begin{lstlisting}[mathescape=true,basicstyle=\ttfamily]
data ExpVal =
    Num_Val {expval_num :: Int}
  | Bool_Val {expval_bool :: Bool}
  | Proc_Val {expval_proc :: Proc}
  | List_Val {expval_list :: [ExpVal]} 
  | Object_Val {expval_obj :: Object}
  | Uninitialized_Val  
        \end{lstlisting} \\
        \bottomrule
    \end{tabular}
    }
    \caption{Different Variants of Expressed Values in the Interpreters}
    \label{tab:expval}
\end{table}

\subsection{Type Definition of \texttt{Store} in \texttt{EXPLICIT-REFS} and \texttt{IMPLICIT-REFS}}

Note that \texttt{Store} is the type for
a mapping of locations to expressed values with a counter to generate fresh locations as \texttt{(Location, [(Location,ExpVal)])}. 

\subsection{The Structure of Interpreters}

The type signatures of various interpreters can illustrate how different language characteristics influence their structure,
as shown in Table \ref{tbl:interp_types}.
For \texttt{LET}, \texttt{PROC}, and \texttt{LETREC}, as well as \texttt{SIMPLE-MODULES}, \texttt{OPAQUE-TYPES}, and \texttt{PROC-MODULES},
the interpreter follows the fundamental type signature \texttt{Exp -> Env -> ExpVal}, 
where expressions are evaluated with an environment to produce a semantic value. 
While this type signature abstracts over differences in environment structures, 
the interpreters for these language features are distinguished by their specific \texttt{Env} definitions, 
which capture the variations in scoping and binding mechanisms, as summarized by Table \ref{tbl:env_definitions} in Appendix.


In \texttt{EXPLICIT-REFS} and \texttt{IMPLICIT-REFS}, an additional \texttt{Store} parameter is introduced
whereas the original EOPL textbook retains the same fundamental type signature thereby making stores implicit. 
This type signature reflects the store-passing style, where evaluation produces a value 
while also updating the store, representing mutable state.
Note that \texttt{Store} is the type for a mapping of locations to expressed values. 
The \texttt{Store} type deserves to appear in the signature because stores are fundamental elements. 
The difference between the two languages can be explained by their differing data types for expressed values. 
In particular, the data type ExpVal for \texttt{EXPLICIT-REFS} includes a variant \texttt{Ref\_Val \{expva\_loc :: Location\}}, 
allowing programs to directly use locations, similar to C, whereas the data type for \texttt{IMPLICT-REFS} does not, as in Java. 
The type definitions of expressed values are available in Table \ref{tab:expval} in Appendix. 

\texttt{CLASSES} and \texttt{TYPED-OO}, inheriting the store-passing style interpreter from \texttt{IMPLICIT-REFS}, 
incorporates \texttt{ClassEnv} to support class-based object-oriented programming.
Their expressed values are ones that extend those of IMPLICIT-REFS with object values and the special value denoting uninitialized fields.



\texttt{LETREC-CPS} and \texttt{EXCEPTIONS} adopt continuation-style interpreters, while \texttt{THREADS} combines this approach with the store-passing style, incorporating \texttt{Cont} to express control contexts. 
They use \texttt{apply\_cont} to apply a continuation to a computed value.
\texttt{EXCEPTIONS} introduces two control flows: a normal execution using \texttt{apply\_cont} and an exceptional flow
using \texttt{apply\_handler}, which handles exceptions (of type \texttt{ExpVal}) by modifying the control flow.
\texttt{THREADS} supports concurrent execution with an arbitrary number of control flows, executing them one by one in a rounding robin schedule. 
In \texttt{THREADS}, while threads in the original EOPL book are represented with a contract that corresponds to the type \texttt{( ) -> ExpVal}, 
our approach instead introduces \texttt{Store -> SchedState -> (ExpVal, Store)}, explicitly managing both store and scheduling state. 
This shift highlights the advantages of using static types, which helped identify an implicit definition of threads in the original EOPL book.

Also, using static types, we identified an imprecise type definition of expressed values 
in which the queue type had to be included as a variant. This was 
due to the structure of storing the runtime-managed thread queue in the store, 
which cannot be directly derived from evaluating \texttt{THREADS} program expressions but
appears in \texttt{ExpVal}
(See Queue\_Val \{expval\_queue :: Queue Thread\} in Table \ref{tab:expval} of the Appendix).

The explicit type signature of \texttt{THREADS} for shared-memory systems proved useful 
in designing an interpreter for a new programming language, \texttt{ACTORS}, for distributed systems, 
as will be discussed in Section \ref{sec:casestudies}.

\section{Type Definitions in the Type Checkers}

\subsection{Type Environment Types}

Table \ref{tab:tyenv} presents different variations of the \texttt{TyEnv} (Type Environment) used in the type checker across different modules. 
Each row corresponds to a specific category of the type checker, with the left column indicating the category and 
the right column showing the corresponding \texttt{TyEnv} definition.
In \texttt{CHECKED} and \texttt{INFERRED}, the type environment consists of an empty environment (\texttt{Empty\_tyenv}) and 
an extension mechanism (\texttt{Extend\_tyenv}) to associate identifiers with types.
In \texttt{SIMPLE-MODULES}, this extends \texttt{TyEnv} by adding \texttt{Extend\_tyenv\_with\_module}, 
allowing modules to be included in the type environment.
In \texttt{OPAQUE-TYPES} and \texttt{PROC-MODULES}, further extends the type environment with \texttt{Extend\_tyenv\_with\_type}, 
which introduces named types.
In \texttt{TYPED-OO}, \texttt{Extend\_tyenv\_with\_self\_and\_super} is introduced, which supports self and supertype relationships, 
useful for class-based type systems.
This table illustrates how the \texttt{TyEnv} structure evolves to accommodate various features in different type checker modules.

\begin{table}[ht]
    \centering
    {\small
    \begin{tabular}{l l}
        \toprule
        Type Checker & Type Environment Definition \\
        \midrule
        \begin{minipage}[t]{.15\linewidth}
        \texttt{CHECKED}, \\ \texttt{INFERRED}
        \end{minipage} &
        \begin{lstlisting}[mathescape=true,basicstyle=\ttfamily]
data TyEnv = Empty_tyenv
  | Extend_tyenv Ident Type TyEnv
        \end{lstlisting} \\
        \midrule
        \texttt{SIMPLE-MODULES} &
        \begin{lstlisting}[mathescape=true,basicstyle=\ttfamily]
data TyEnv = Empty_tyenv
  | Extend_tyenv Ident Type TyEnv
  | Extend_tyenv_with_module Ident Interface TyEnv
        \end{lstlisting} \\
        \midrule
        \begin{minipage}[t]{.15\linewidth}
        \texttt{OPAQUE-TYPES}, \\ \texttt{PROC-MODULES}
        \end{minipage}
        &
        \begin{lstlisting}[mathescape=true,basicstyle=\ttfamily]
data TyEnv = Empty_tyenv
  | Extend_tyenv Ident Type TyEnv
  | Extend_tyenv_with_module Ident Interface TyEnv
  | Extend_tyenv_with_type Ident Type TyEnv
        \end{lstlisting} \\
        \midrule
        \texttt{TYPED-OO} &
        \begin{lstlisting}[mathescape=true,basicstyle=\ttfamily]
data TyEnv = Empty_tyenv
  | Extend_tyenv Ident Type TyEnv
  | Extend_tyenv_with_self_and_super Type Ident TyEnv
        \end{lstlisting} \\
        \bottomrule
    \end{tabular}
    }
    \caption{Different Variants of \texttt{TyEnv} in the Type Checker}
    \label{tab:tyenv}
\end{table}

\subsection{Type and Interface Definitions}

Table \ref{tab:type} presents the different variations of the \texttt{Type} definition used in various type checkers. Each row corresponds to a specific category of the type checker, with the left column indicating the category and the right column showing the corresponding \texttt{Type} definition.
In \texttt{CHECKED} and \texttt{SIMPLE-MODULES}, the basic type system consists of primitive types \texttt{TyInt} and \texttt{TyBool}, along with function types represented by \texttt{TyFun}.
In \texttt{INFERRED}, this extends the type system by adding \texttt{TyVar}, which represents type variables identified by integers, enabling type inference.
\texttt{OPAQUE-TYPES} and \texttt{PROC-MODULES} introduce \texttt{TyName} and \texttt{TyQualified}, which allow named and qualified types, respectively, enhancing modularity and abstraction.
\texttt{TYPED-OO} further expand the type system with \texttt{TyVoid} (representing void types), \texttt{TyFun} supporting lists of argument types, \texttt{TyClass} for class-based types, and \texttt{TyListOf} for list types.
This table illustrates how the type system evolves to accommodate different levels of abstraction and modularity in various type checking environments.

Table \ref{tab:interface}  outlines the variations of the \texttt{Interface} definition across different type checkers. The left column specifies the type checker category, while the right column contains the corresponding \texttt{Interface} definition.
In \texttt{SIMPLE-MODULES} and \texttt{OPAQUE-TYPES}, we define a basic module interface with \texttt{SimpleIface}, which consists of a list of declarations.
\texttt{PROC-MODULES} extends the interface definition by introducing \texttt{ProcIface}, which allows for procedural module interfaces. This enables structured interaction between different modules, supporting hierarchical compositions of interfaces.
This table highlights the evolution of interface definitions, demonstrating how more advanced module composition features are incorporated in different type checkers.

\begin{table}[ht]
    \centering
    {\small
    \begin{tabular}{l l}
        \toprule
        Type Checker & Type Definition \\
        \midrule
        \begin{minipage}[t]{.15\linewidth}
        \texttt{CHECKED}, \\ \texttt{SIMPLE-MODULES}
        \end{minipage}
        &
        \begin{lstlisting}[mathescape=true,basicstyle=\ttfamily]
data Type = TyInt | TyBool | TyFun Type Type
        \end{lstlisting} \\
        \midrule
        \texttt{INFERRED} &
        \begin{lstlisting}[mathescape=true,basicstyle=\ttfamily]
data Type = TyInt | TyBool | TyFun Type Type 
          | TyVar Integer
        \end{lstlisting} \\
        \midrule
        \begin{minipage}[t]{.15\linewidth}
        \texttt{OPAQUE-TYPES}, \\ \texttt{PROC-MODULES}
        \end{minipage} &
        \begin{lstlisting}[mathescape=true,basicstyle=\ttfamily]
data Type = TyInt | TyBool | TyFun Type Type 
          | TyName Ident 
          | TyQualified Ident Ident
        \end{lstlisting} \\
        \midrule
        \texttt{TYPED-OO} &
        \begin{lstlisting}[mathescape=true,basicstyle=\ttfamily]
data Type = TyInt | TyBool | TyVoid | TyFun [Type] Type 
          | TyClass Ident | TyListOf Type
        \end{lstlisting} \\
        \bottomrule
    \end{tabular}
    }
    \caption{Different Variants of Type in the Type Checker}
    \label{tab:type}
\end{table}

\begin{table}[ht]
    \centering
    {\small
    \begin{tabular}{l l}
        \toprule
        Type Checker & Interface Definition \\
        \midrule
        \begin{minipage}[t]{.3\linewidth}
        \texttt{SIMPLE-MODULES}, \texttt{OPAQUE-TYPES}
        \end{minipage}
        &
        \begin{lstlisting}[mathescape=true,basicstyle=\ttfamily]
data Interface = SimpleIface [ Declaration ]
        \end{lstlisting} \\
        \midrule
        \texttt{PROC-MODULES} &
        \begin{lstlisting}[mathescape=true,basicstyle=\ttfamily]
data Interface = SimpleIface [ Declaration ]
  | ProcIface Ident Interface Interface
        \end{lstlisting} \\
        \bottomrule
    \end{tabular}
    }
    \caption{Different Variants of Interface in the Type Checker}
    \label{tab:interface}
\end{table}  

\subsection{The Structure of Type Checkers}

The type signatures of various type checkers also illustrate their fundamental structure and variations,
as shown in Table \ref{tbl:typechecker_types}.
The basic type-checking structure follows the signature \texttt{Exp -> TyEnv -> Either ErrMsg Type}, 
where an expression (\texttt{Exp}) and a type environment (\texttt{TyEnv}) are given as input,
returning either a successfully inferred type (\texttt{Type}) or an error message (\texttt{ErrMsg}) if the expression is ill-typed.
This structure is adopted by the type checkers in \texttt{CHECKED} and \texttt{SIMPLE-MODULES}.
\texttt{TYPED-OO} introduces a static class environment (\texttt{StaticClassEnv}) to the basic signature,
ensuring that type checking considers class hierarchies.

\texttt{INFERRED} allows programmers to omit type declarations in their programs, with a type inference system automatically determining 
and filling in the missing types.  The system extends the type-checking process by generating fresh type variables for missing types
and solving type equations constrained by these variables. 
The type signature of the type inference in \texttt{INFERRED}
introduces a type substitution (\texttt{Subst}) parameter 
and replaces \texttt{Either} with \texttt{Either\_}, a variant that incorporates an internal counter state. 
The first change reflects how type inference proceeds by analyzing the structure of \texttt{Exp},
gradually refining the input type substitution to include more constraints on type variables as it progresses.
For example, in the expression \texttt{Call\_Exp rator rand}, after inferring the types \texttt{ty1} and \texttt{ty2} 
for \texttt{rator} and \texttt{rand}, respectively, the system unifies \texttt{(ty1, ty2 -> fresh\_tyvar)}. At this point,
an internal count is incremented to generate the fresh type variable, and upon successful unification, the system returns it. 

The second change accounts for both side effects: determining whether type inference succeeds or fails (\texttt{ExceptT e m a}) 
and managing a counter state for generating new type variable names during the inference process (\texttt{State Integer} for \texttt{m}).
\texttt{Either\_ e a} is defined as \texttt{type  Either\_ e a = ExcetpT e m a}
which represents a monadic computation that combines 
exception handling (\texttt{ExceptT}) and state management (\texttt{State Integer}) \cite{wadler1992} \cite{sheng1995}.
This definition extends error handling with stateful computations, enabling fresh name generation,
while maintaining nearly the same structure as when \texttt{Either} was used.
We deliberately designed a custom-defined data type, \texttt{Either\_}, to resemble \texttt{Either}, which is  
a builtin algebraic data type defined as \texttt{data Either a b = Left a | Right b}, and represents 
computations that may result in either success or failure. 
%
An advantage of using \texttt{Either} and \texttt{Either\_} is that they explicitly convey the scope of side effects in the alternative and stateful computations within the type checker implementations. 

In \texttt{OPAQUE-TYPES} and \texttt{PROC-MODULES}, the type checkers handle opaque types and procedural modules using essentially the same type signature as in \texttt{CHECKED} and \texttt{SIMPLE-MODULES}, namely \texttt{Exp -> TyEnv -> Either\_ ErrMsg Type}. However, they rely on $\texttt{Either\_},\ldots$ rather than $\texttt{Either},\ldots$.
In \texttt{OPAQUE-TYPES}, fresh module names must be generated to avoid conflicts when opaque types are compared with qualified types, thereby keeping each interface clearly separated from its implementation. In \texttt{PROC-MODULES}, fresh module names are also needed for argument modules when checking the compatibility of module procedure interfaces.
These requirements for fresh names are reflected in the use of the variant type with $\texttt{Either\_}$.

Note that since \texttt{Store} and \texttt{Subst} are essential components of programming languages, 
they deserve to be explicitly included in the type signature.
In contrast, the counter state for generating new names is not a major component, so 
its existence was merely highlighted through the \texttt{Either\_} type, thus making it implicit.

\end{document}